\documentclass[twocolumn,showpacs,floats,floatfix,superscriptaddress,aps,pra]{revtex4-1}
\usepackage{amsfonts}
\usepackage{amssymb}
\usepackage{amsmath}
\usepackage{calc}
\usepackage{graphicx}
\usepackage{bm}

\usepackage[normalem]{ulem}
\usepackage{amsmath,amssymb}
\usepackage{multirow}
\usepackage{xcolor,soul}
\usepackage{srcltx}
\usepackage{hyperref,graphicx}

\begin{document}

\author{Gian Luca Giorgi}
\affiliation{ IFISC (UIB-CSIC), Instituto de Fisica Interdisciplinar y Sistemas Complejos - Palma de Mallorca, E-07122. Spain}
\author{ Salvatore Lorenzo}  
\affiliation{Dipartimento di Fisica e Chimica, Universit\'a degli Studi di Palermo, via Archirafi 36, I-90123 Palermo, Italy}
\author{Stefano Longhi}
\affiliation{ Dipartimento di Fisica, Politecnico di Milano and Istituto di Fotonica e Nanotecnologie del Consiglio Nazionale delle Ricerche, Piazza L. da Vinci 32, I-20133 Milano, Italy}
\affiliation{ IFISC (UIB-CSIC), Instituto de Fisica Interdisciplinar y Sistemas Complejos - Palma de Mallorca, E-07122. Spain}
\email{stefano.longhi@polimi.it}

\title{Topological Protection and Control of Quantum Markovianity}
  \normalsize

%\date{.}

%
\bigskip
\begin{abstract}
\noindent  
Under the Born--Markov approximation, a qubit system, such as a two-level atom,  is known to undergo a memoryless decay of quantum coherence or excitation when weakly coupled to a featureless environment.  Recently, it has been shown that unavoidable disorder in the environment is responsible for non-Markovian effects and information backflow from the environment into the system owing to Anderson localization. This~turns disorder into a resource for enhancing non-Markovianity in the system--environment dynamics, which could be of relevance in cavity quantum electrodynamics. Here we consider the decoherence dynamics of a qubit weakly coupled to a two-dimensional bath with a nontrivial topological phase, such as a two-level atom embedded in a two-dimensional coupled-cavity array with a synthetic gauge field realizing a quantum-Hall bath, and~show that Markovianity is protected against moderate disorder owing to the robustness of chiral edge modes in the quantum-Hall bath. Interestingly, switching off the gauge field, i.e., flipping the bath into a topological trivial phase, allows one to re-introduce non-Markovian effects. Such a result indicates that changing the topological phase of a bath by a tunable synthetic gauge field can be harnessed to control non-Markovian effects and quantum information backflow in a qubit-environment system.
\end{abstract}

\maketitle

\section{Introduction}
Relaxation and decoherence dynamics in open quantum systems is attracting a continuous and increasing interest since more than three decades, with
major relevance both in foundations of quantum physics, such as for the explanation of spontaneous quantum decay and 
the quantum to classical transition \cite{deco1}, as~well as for a wide variety of physical problems ranging from quantum engineering to many-body systems and quantum information science \cite{deco2,deco3}, where 
 irreversible dynamical behaviors such as energy dissipation, relaxation to a thermal equilibrium, and~decay of quantum coherence and correlations are commonplace. In the majority of cases, where system and environment time scales are widely separated, the~evolution of the reduced density matrix of
a quantum system weakly coupled to a featureless environment is governed by a master equation of the Lindblad form, which describes a memoryless dynamics
typically leading to an irreversible loss of quantum features. However, with recent advances in quantum technologies 
and quantum engineering  memory effects are becoming experimentally relevant. Therefore, great interest is currently devoted to the search and control of memory effects in large environments, as~well as the development of theoretical  tools to quantify the amount of non-Markovianity \cite{deco4,deco5,deco6,deco7}. Strong system--environment coupling, edge effects and structured reservoirs are the most  common causes of memory effects
and revivals of genuine quantum properties such as quantum coherence and entanglement \cite{deco4,deco5,deco6}. In some recent works, it has been suggested and demonstrated that disorder in the bath 
can be exploited to realize strong coupling conditions  for light--matter interaction \cite{An1,An2} and to enhance non-Markovian effects \cite{An3,An4} owing to Anderson localization \cite{An5}. Likewise, memory effects driven by a metal-insulator phase transition has been predicted to occur in a bath described by an incommensurate quasi periodic potential \cite{An6}, and~the impact of long-range disorder on the creation and distribution of entanglement has been investigated as well in a spin chain model \cite{An7}. The~interplay between many-body interaction and localization, quantum phase transitions and non-Markovianity has been theoretically investigated as well in some recent works \cite{An8,An9,An10,An11}. Most of previous studies on memory effects in open quantum systems deal rather generally with 
a topologically trivial environment. The~interplay between topological order, quantum decay and non-Markovianity is a largely unexplored area of research, which is just being considered in some recent works \cite{NMT1,NMT2,NMT3,NMT4,NMT5,NMT6}. 
Topological phases of matter are the most active research fields in modern condensed matter physics, photonics and several other areas of physics \cite{T1,T2,T3,T4,T5,T6,T7,T8,T9,T10,T11,T12,T13,T14,T15,T16,T17}, and~the ability of engineering 
a quantum bath to be topologically nontrivial is expected to become of experimental relevance in the near future. Some nontrivial features are expected to arise when the excitation decay or decoherence occur into a topologically nontrivial continuum \cite{An9,An10,An11}. The~interplay between disorder and topological order is a subject of great interest \cite{disor0,disor1,disor2,disor3,disor4,disor5,disor6,disor7,disor8}. In particular, chiral edge states in two- and three-dimensional topological insulators are known to be rather generally robust against mild to moderate disorder in the system \cite{disor0}, while quite remarkably disorder can introduce non-trivial topological phases in systems that are topologically trivial in the crystalline phase (i.e., without disorder), leading to so-called topological Anderson insulators \cite{disor5,disor6,disor7,disor8}.

In this work we investigate the  decoherence and non-Markovian dynamics of a qubit interacting with a quantum bath with reconfigurable topological phase and with disorder. The~main result of our analysis is that, while in a topologically-trivial quantum environment disorder induces strong memory effects owing to Anderson localization \cite{An3}, flipping the topological phase of the bath into a non-trivial protects Markovianity against disorder in the system. The~main idea is illustrated by considering the decoherence of a two-level atom embedded in an edge resonator of a two-dimensional array of coupled optical cavities with a reconfigurable synthetic gauge field, which realizes a two-dimensional quantum Hall environment \cite{T1}. Our results indicate that reconfigurable topological baths in connection with disorder can be exploited to controlling (either suppressing or enhancing) memory effects in open quantum systems.

%%%%%%%%%%%%%%%%%%%%%%%%%%%%%%%%%%%%%%%%%%
\section{Model and Decoherence Dynamics}
\begin{figure}%[htbp]
  \includegraphics[width=85mm]{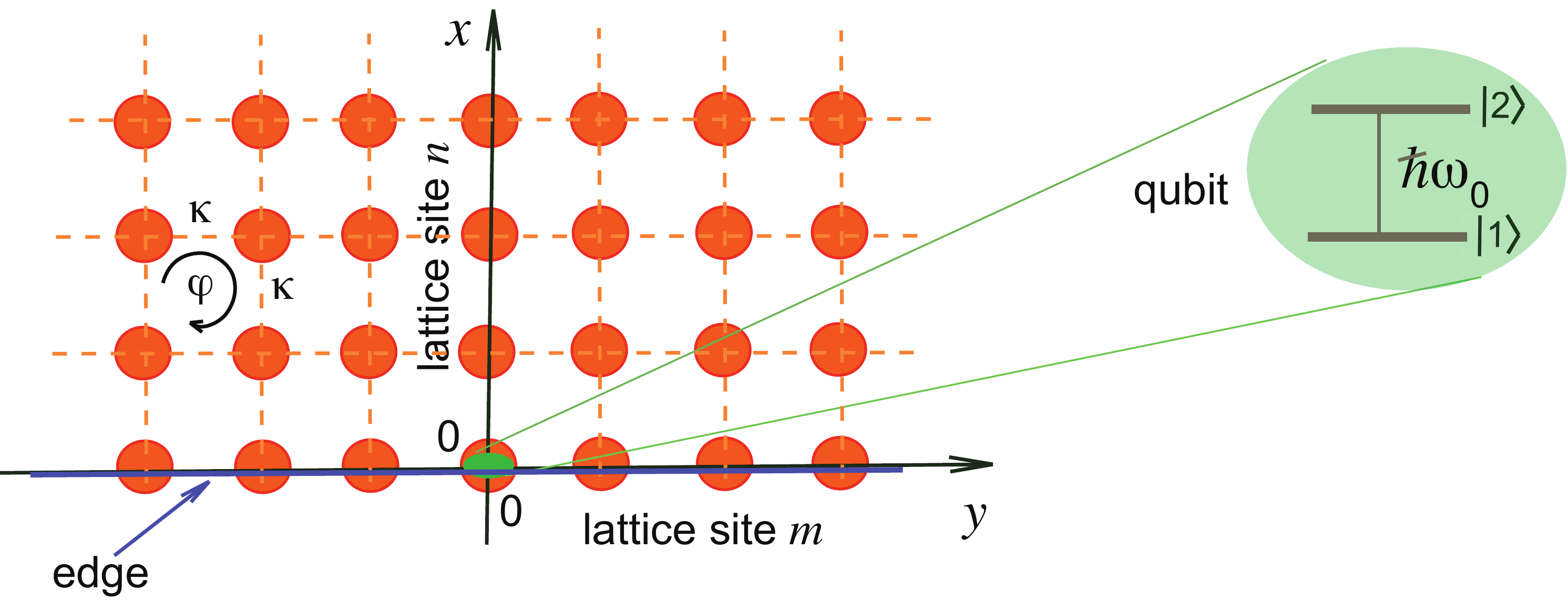}\\
   \caption{(color online) Schematic of a two-dimensional semi-infinite square lattice of optical resonators with resonance frequencies $\omega_c+\delta \omega_{n,m}$, nearest-neighbor evanescent field coupling $\kappa$, and~artificial gauge field $\varphi$. A~two-level atom (qubit) with resonance frequency $\omega_0$ close to $\omega_c$ is embedded in one of edge resonators at site $n=m=0$. The~array is infinitely extended in the $y$ direction. For~a non-vanishing gauge field $\varphi$, the~coupled-resonator array realizes a topological nontrivial bath interacting with the~qubit.}
   \label{Fig1} 
\end{figure}

We consider a semi-infinite two-dimensional square array of coupled optical resonators and a qubit, such as a two-level atom, embedded in a boundary resonator of the array, as~schematically shown in Figure~\ref{Fig1}. A~synthetic gauge field is applied to the square lattice resonator so as to introduce Peierls$^{\prime}$ phases and a non-vanishing effective magnetic flux $\varphi$ in each plaquette of the array, thus realizing a photonic quantum-Hall topological insulator with chiral edge modes. The~artificial gauge field can be accomplished, for instance, by dynamic modulation, using auxiliary cavities, or by other means, as~discussed and demonstrated in several recent works \cite{T13,topo0,topo1,topo2,topo3,topo4,topo5,topo6,topo6b,topo7,topo7c}. We also mention that different types of reconfigurable two-dimensional photonic lattice configurations, with controllable switching between topologically trivial and non-trivial phases, have been suggested and demonstrated in several other different photonic setups \cite{topo8,topo9,topo10,topo11,topo12,topo13}.

The bath into which the qubit is coupled is thus represented by a two-dimensional quantum-Hall (Harper--Hofstadter) system, which shows a non-trivial topological phase and chiral edge states for rational values of the non-vanishing magnetic flux.  The Hamiltonian of the system (with $\hbar=1$) reads 
\begin{equation}
\hat{H}=\hat{H}_{p}+\hat{H}_{bath}+\hat{H}_I
\end{equation}
where
\begin{equation}
\hat{H}_p= \omega_0 \sigma_{+}\sigma_{-}
\end{equation}
is the Hamiltonian of the two-level atom with resonance frequency $\omega_0$, $\sigma_{\pm}$ are the raising and lowering operators of ground ($| g \rangle$) and excited ($|e \rangle$) levels,
\begin{eqnarray}
\hat{H}_{bath} & = &  \sum_{n,m} (\omega_c+ \delta \omega_{n,m} )  \hat{a}^{\dag}_{n,m} \hat{a}_{n,m}+ \\
& + & \kappa \sum_{n,m} \left\{ \hat{a}^{\dag}_{n,m} \hat{a}_{n+1,m}+\hat{a}^{\dag}_{n,m} \hat{a}_{n,m+1} \exp(2 \pi in \varphi) +H.c. \right\} \nonumber
\end{eqnarray}
is the Hamiltonian of the photon field in the resonator array (coupling constant $\kappa$, cavity resonance frequency $\omega_c$,  and synthetic magnetic flux $\varphi$), and
\begin{equation}
\hat{H}_I= \Delta \left( \sigma_+ \hat{a}_{0,0}+\sigma_- \hat{a}_{0,0}^{\dag} \right)
\end{equation}
is the interaction Hamiltonian in the rotating-wave approximation. In the above equations, $\hat{a}_{n,m}^{\dag}$ and  $\hat{a}_{n,m}$ are the creation and destruction operators of the single-mode photon field in the $(n,m)$ resonator of the semi-infinite array, with $m=0, \pm1, \pm 2, \pm 3, ...$ and $n=0,1,2,3,...$, $\delta \omega_{n,m} $ describes some deviations (disorder) of the cavity resonance frequency of resonator $(n,m)$ from the reference value $\omega_c$,
$\Delta$ is the atom--photon coupling strength, and~the two-level atom is embedded in the $(n=0,m=0$) edge resonator of the array (Figure~\ref{Fig1}). We assume that at initial time $t=0$ the atom is prepared in the excited state while the photon field in the square lattice is in the vacuum state, i.e., $|\psi(0) \rangle= |e \rangle \otimes |0 \rangle$. Apart from a phase term rotating at the frequency $\omega_0$, the~evolved state at time $t$ is given by
\begin{equation}
| \psi(t) \rangle= q(t) |e \rangle \otimes |0 \rangle + | g \rangle \otimes \left( \sum_{n,m} \alpha_{n,m}(t) \hat{a}^{\dag}_{n,m} | 0 \rangle \right)
\end{equation}
where the amplitudes $q(t)$ and $\alpha_{n,m}$ satisfy the coupled equations
\begin{eqnarray}
i \frac{dq}{dt} & = & \Omega q+\Delta \alpha_{0,0} \\
i \frac{d \alpha_{n,m}}{dt} & = & \mathcal{H}^{(QH)} \alpha_{n,m}+ \Delta \delta_{n,0} \delta_{m,0}  q
\end{eqnarray}
($m=0, \pm 1 , \pm 2, ...$, $n=0,1,2,3,...$) with the initial condition 
\begin{equation}
q(0)=1, \; \; \alpha_{n,m}(0)=0.
\end{equation}
In Equations~(6) and (7), $\mathcal{H}^{(QH)}$ is the Hamiltonian of the quantum-Hall bath, i.e.,
\begin{widetext}
\begin{equation}
\mathcal{H}^{(QH)} \alpha_{n,m}=  \delta \omega_{n,m} \alpha_{n,m}+\kappa \left\{ \alpha_{n+1,m}+ \alpha_{n-1,m}+ \exp(2 \pi i n \varphi ) \alpha_{n,m+1}+ \exp(-i n \varphi) \alpha_{n,m-1} \right\}.
\end{equation}
\end{widetext}
and
\begin{equation}
\Omega \equiv  \omega_0-\omega_c
\end{equation}
is the frequency detuning between the resonance frequency $\omega_0$ of the two-level atom and the reference frequency $\omega_c$ of the optical mode of the resonators in the array. Clearly, $|q(t)|^2$ is the probability that the two-level system remains in the excited state at time $t$. More generally, if at initial time the qubit is in a mixed state described by the density matrix $\rho_{i,k}=\langle i | \rho(0) | k \rangle$ ($i,k=g,e$) and the photon field in the resonator lattice is in the vacuum state, the~reduced density matrix $\rho(t)$  of the qubit at time $t$, obtained after tracing out the lattice degrees of freedom, reads
  \begin{equation}
  \rho(t)= \left(
  \begin{array}{cc}
  |q(t)|^2 \rho_{ee} & q(t) \rho_{eg} \\
  q^*(t) \rho_{ge} & (1-|q(t)|^2) \rho_{ee}+\rho_{gg}
  \end{array}
  \right).
  \end{equation}
 Therefore, the~coherence of the qubit decays as $\sim |q(t)|$, where $q(t)$ is the solutions to Equations~(6) and (7) with the initial conditions defined by Equation~(8).
Oscillations in the decay behavior of $q(t)$  correspond to a periodic reflux of information from the bath back to the qubit, which can be studied in terms of a non-Markovianity quantifier $\mathcal{N}$ \cite{qu1,qu2} . Such a quantifier of memory flow-back is based on the observation that in  Markovian processes the distinguishability between pairs of quantum
states decreases in time, and~can be defined  as the integral of $d |q(t)|/dt$ extended over the time intervals where the derivative $\dot |q(t)|$ is positive. As discussed in previous works \cite{An3,NMT6}, persistent oscillations in the behavior of $q(t)$, such those arising from Anderson localization in the continuum bath, would lead to a divergence of the non-Markoviantity quantifier. A~simple regularization procedure is to introduce the temporal average \cite{NMT6}
 \begin{equation}
{\mathcal N}_T= \frac{1}{T} \int_{0 \; \dot{|q|}>0}^{T} dt \frac{d 
|q(t)|}{dt},
\end{equation}
where the integral is extended over the time intervals where the derivative  { $(d|q|/dt)$} is positive.

%%%%%%%%%%%%%%%%%%%%%%%%%%%%%%%%%%%%%%%%%%
\section{Control of Quantum Non-Markovianity by a Gauge Field}

\begin{figure*}%[htbp]
  \includegraphics[width=160mm]{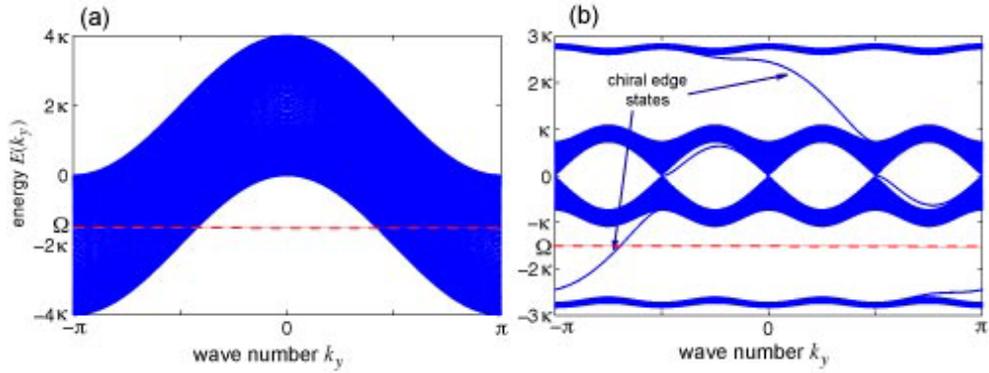}\\
   \caption{(color online)  Energy dispersion curve $E=E(k_y)$ of the Harper--Hofstadter eigenvalue Equation (14) versus the wave number $k_y$ in the translational invariant direction $y$ for (\textbf{a}) $\varphi=0$, corresponding to a topologically trivial bath, and~(\textbf{b}) $\varphi=p/q$ with $p=1$ and $q=4$, corresponding to a topologically nontrivial bath. In (\textbf{b}) the single tight-binding band of (\textbf{a}) splits into $q=4$ magnetic narrow bands with non vanishing Chern numbers, separated by three gaps. In the upper and lower wide gaps chiral edge states, with opposite propagation directions, are clearly observed. The~dashed horizontal line denotes the frequency detuning $\Omega$ of the qubit. }
   \label{Fig2} 
\end{figure*}

The decoherence dynamics of the quit and the non-Markovianity quantifier $\mathcal{N}_T$ are determined by the decay behavior of $q(t)$, which requires to numerically compute the solution to the infinite- dimensional coupled Equations (6) and (7) with the initial condition (8). In this section we unveil the major impact of the synthetic gauge field on the decoherence dynamics in the presence of disorder in the bath. 
 Before discussing the main numerical results, it is useful to provide a simple physical explanation of the impact of the gauge field on the decay behavior of $q(t)$. To this aim, let us first assume that there is not any disorder in the bath, i.e., $\delta \omega_{n,m}=0$ in Equation~(9). In this case, owing to the translation invariance of the bath along the $y$ direction, the~scattering states $\psi_{n,m}$ of the Hamiltonian $\mathcal{H}^{(QH)}$, satisfying the eigenvalue equation $\mathcal{H}^{(QH)}\psi_{n,m}=E(k_y) \psi_{n,m}$, are of the form
 \begin{equation}
 \psi_{n,m}=A_n \exp(ik_ym)
 \end{equation}
{($- \pi \leq k_y < \pi$) where $A_n$ and the eigen-energy $E(k_y)$ satisfy the famous Harper--Hofstadter equation~\cite{harper,hof}}
 \begin{equation}
  \kappa(A_{n+1}+A_{n-1})+2 \kappa \cos (2 \pi \varphi n +k_y) A_n=E(k_y)A_n
 \end{equation}
with $n=0,1,2,3,..$ and with the boundary condition $A_{-1}=0$. For~$\varphi=0$, the~eigenstates of Equation~(14) are scattering states, given by $A_n= \sin [(n+1)k_x]$, with $ -\pi \leq k_x < \pi$, while there are not any bund state localized near the edge $n=0$. In this case the energy spectrum of $\mathcal{H}^{(QH)}$ is gapless and defined by the simple dispersion relation 
\begin{equation}
E(k_x,k_y)=2 \kappa \cos (k_x)+ 2 \kappa \cos (k_y)
\end{equation}
showing a single band of width $8 \kappa$ as the wave number $k_y$ varies from $-\pi$ to $\pi$ (Figure~\ref{Fig2}a). By varying the phase $\varphi$ the energy spectrum $E(k_y)$ shows a fractal structure, known as the Hofstadter butterfly \cite{hof}. 
For $\varphi \neq 0$, the~model explicitly breaks time-reversal symmetry because of the Peierls phase factors, which leads to the appearance of topologically non-trivial energy bands, i.e., bands characterized by a non-vanishing Chern number \cite{T1}. For~rational values of $\varphi=p/q$, the~band (15) splits into $q$ magnetic sub bands with a wide family of topologically-protected quantum Hall chiral edge states~\cite{T1}. For~example, for $p=1$ and $q=4$, the~energy band (15) splits into four magnetic sub bands, with two wide gaps sustaining topologically-protected chiral edge states propagating in opposite directions, as~shown in Figure~\ref{Fig2}b.

\begin{figure}%[htbp]
  \includegraphics[width=85mm]{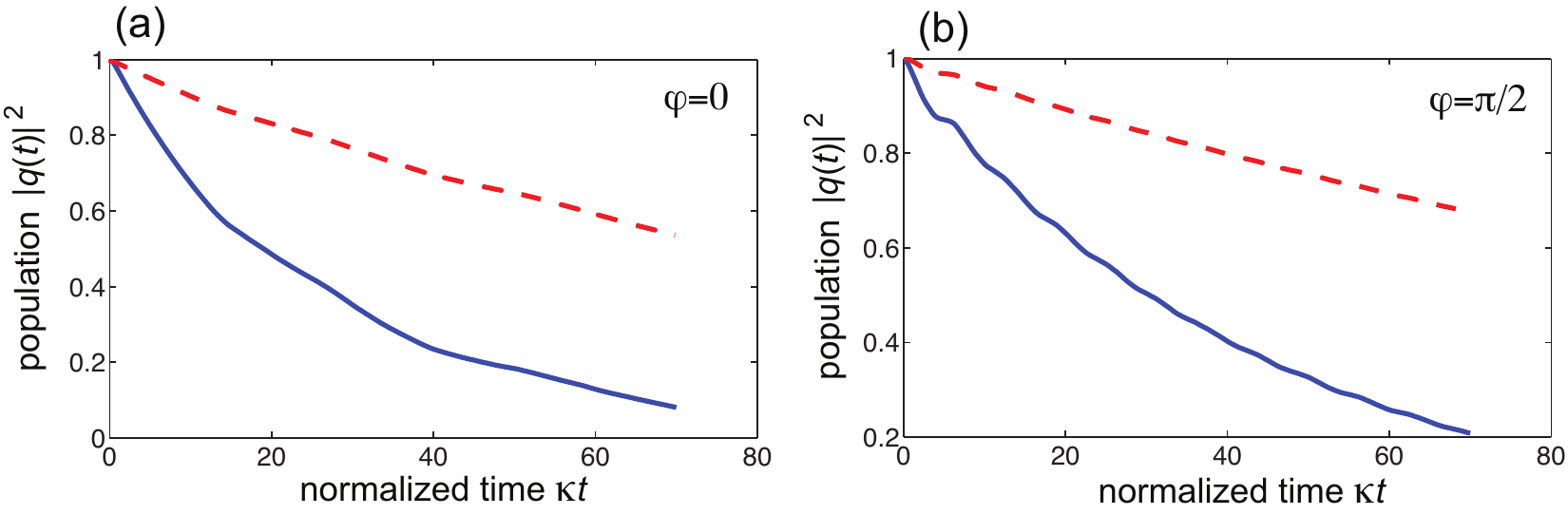}\\
   \caption{(color online) Decay dynamics of the two-level atom in a two-dimensional bath (behavior of $|q(t)|^2$ versus normalized time $\kappa t$) in the absence of disorder for (\textbf{a}) $\varphi=0$ (topologically-trivial bath), and~(\textbf{b}) $\varphi= \pi/2$ (quantum-Hall bath). Solid and dashed curves refer to the normalized atom-field couplings $\Delta / \kappa=0.2$ and $\Delta / \kappa=0.1$, respectively. Detuning is set at $\Omega / \kappa=-3/2 $. }
   \label{Fig3} 
\end{figure}

\begin{figure}%[htbp]
  \includegraphics[width=85mm]{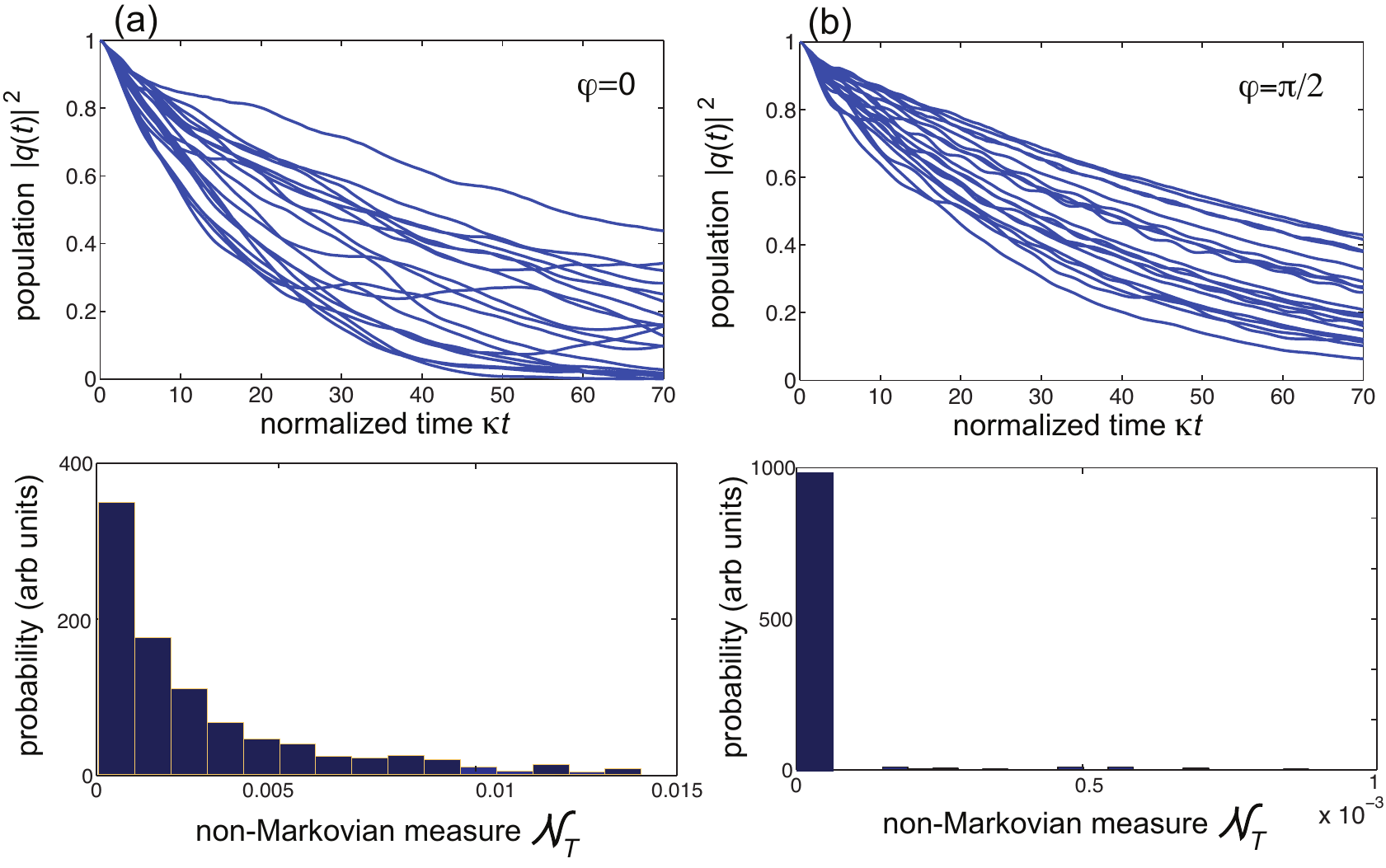}\\
   \caption{(color online) Effect of disorder on the decay dynamics and non-Markovianity quantifier $\mathcal{N}_T$ in (\textbf{a}) the topologically-trivial bath $(\varphi=0)$, and~(\textbf{b}) in the quantum-Hall bath ($\varphi= \pi/2$).
The upper panels show the numerically-computed behavior of the upper-level population $|q(t)|^2$ versus normalized time $\kappa t$ for 20 different realizations of disorder of resonator resonance frequencies. $\delta \omega_{n,m}$ is assumed to be a random number with zero mean and uniform distribution in the interval  $(-\delta,\delta)$, where $\delta= \kappa$ is the disorder strength. Other parameter values are $\Delta / \kappa=0.2$ and $\Omega / \kappa=-3/2$. The~lower plots show the numerically-computed probability distribution of the non-Markovian quantifier $\mathcal{N}_T$ as obtained for 1000 different realizations of disorder.  Note that in the topologically non-trivial bath sustaining chiral edge states [histogram in panel (\textbf{b})] the probability distribution is squeezed toward $\mathcal{N}_T=0$, indicating a high degree of Markovianity protection}
   \label{Fig4} 
\end{figure}

To unveil the very different decay behavior in the presence of disorder for the topologically-trivial bath $\varphi=0$ of Figure~\ref{Fig2}a and the topologically non-trivial bath $\varphi= \pi/2$ of Figure~\ref{Fig2}b, let us assume that the frequency detuning $\Omega$ of the two-level atom [Equation~(10)] falls inside a wide magnetic gap of the topological quantum Hall bath (see the horizontal dashed curves in Figure~\ref{Fig2}). Clearly, in the $\varphi=0$ case the two-level atom decay arises from the coupling with the two-dimensional scattering states of the bath at frequency in the band close to $\Omega$, while for $\varphi=\pi/2$ the decay arises from the coupling with the chiral edge states of the bath (rather than with bulk states). For~a weak coupling $\Delta \ll \kappa$ and in the absence of disorder, the~decoherence dynamics is Markovian for both $\varphi=0$ and $\varphi= \pi/2$, as~shown in Figure~\ref{Fig3}. The~slightly faster decay rate in the topological trivial case $\varphi=0$ is basically due to a larger density of states at frequency $\Omega$, and~the decay law becomes closer to an exponential law, with a decay rate given by the Fermi golden rule, as~the coupling atom-field coupling $\Delta$ becomes smaller. Since disorder acts very different for bulk end edge chiral states, we expect a very distinct behavior in the decoherence dynamics for the disordered lattice when tuning the magnetic flux from $\varphi=0$ to $\varphi= \pi/2$.  This is illustrated in Figure~\ref{Fig4}, where the detuning $\delta \omega_{n,m}$ of resonance frequencies is assumed to be a random number with zero mean and uniform distribution in the interval  $(-\delta,\delta)$. For~a lattice with weak-to-moderate disorder ($| \delta \omega_{n,m}|$ smaller than $\sim \kappa$), Anderson localization arises in the bulk states of the topologically-trivial bath of Figure~\ref{Fig2}a, thus leading to noticeable non-Markovian dynamics because of Rabi-like flopping as previously shown in Ref. \cite{An3}. This~is clearly shown in Figure~\ref{Fig4}a, which depicts the numerically-computed behavior of the population decay laws $|q(t)|^2$ and corresponding values of the non-Markovian quantifier $\mathcal{N}_T$ for 20 different realizations of disorder. Note the dependence of the decay dynamics on the disorder realization and the appearance of oscillations. On the other hand, when the magnetic flux is tuned to $\varphi/2$, the~chiral edge states are robust against localization and the Markovianity of decoherence dynamics is protected, as~shown in Figure~\ref{Fig4}b. The~lower panels in Figure~\ref{Fig4} show the statistical distribution of the non-Markovian quantifier, for $\varphi=0$ and $\varphi= \pi/2$, as~obtained by considering 1000 different realizations of disorder in the lattice. Clearly, in the topologically non-trivial phase the statistical distribution is almost fully squeezed toward $\mathcal{N}_T=0$, indicating that with probability higher than $\sim 98 \%$ the decay remains Markovian despite the disorder. The~rare event where decay becomes non-Markovian in the $\varphi= \pi/2$ phase occurs when the disorder is strong in the neighborhood of the resonator $n=m=0$ of the lattice, which locally perturbs the chiral state with energy flow back into the two-level atom, while disorder in resonator frequencies far from the $n=m=0$ resonator does not substantially alter the decay dynamics. On the other hand, in the trivial phase $\varphi=0$ disorder induces localization of bulk eigenmodes of the 2D lattice and non-Markovianity is enhanced because of Rabi-like flopping dynamics between the qubit and near-resonant localized bulk modes. We note that the same scenario of disorder-induced non-Markovianity would be observed if the qubit were coupled to a topologically-trivial 1D  tight-binding lattice with non-chiral propagating modes, as~previously shown in \cite{An3}.

It should be also emphasized that, for the observation of topological protection of Markovianity, it is crucial that the qubit is primarily coupled to the chiral edge states of the quantum Hall bath in the non-trivial regime, i.e., that $\Omega$ falls inside a magnetic gap of the quantum Hall insulator (Figure~\ref{Fig2}b). Indeed, if the qubit were significantly coupled to bulk modes, one would observe a less pronounced difference between the two phases.

\begin{figure}
\includegraphics[width=8.6 cm]{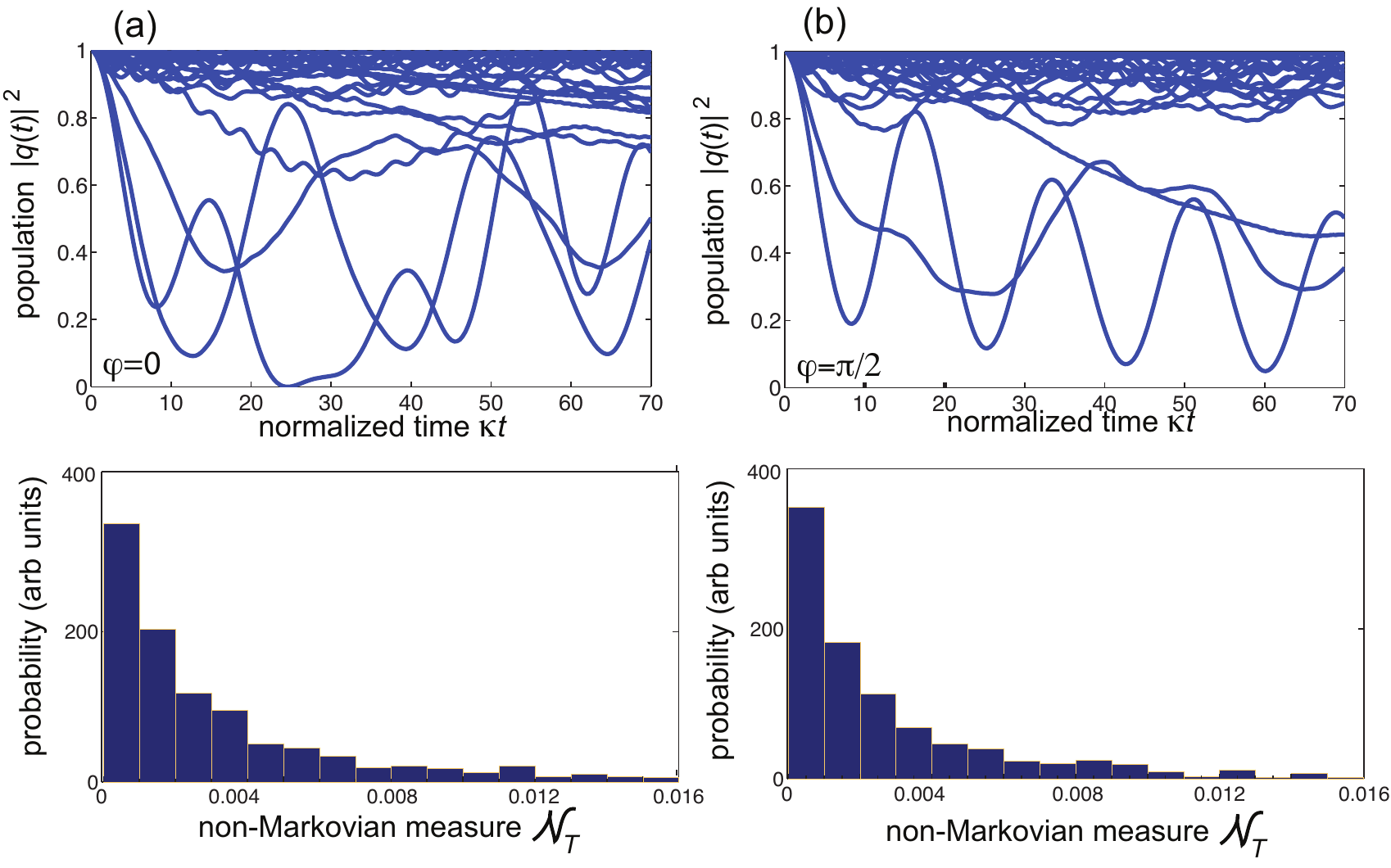}
\caption{{{Same as Figure}~\ref{Fig4}, but in the strong disorder regime ($\delta= 5 \kappa$). In this case the quantum Hall bath becomes an Anderson insulator and the topological protection of Markovianity is not observed~anymore.}}%There is no explanation for the subfigure in the figure.
\label{Fig5}
\end{figure}

\begin{figure}
\includegraphics[width=8.4 cm]{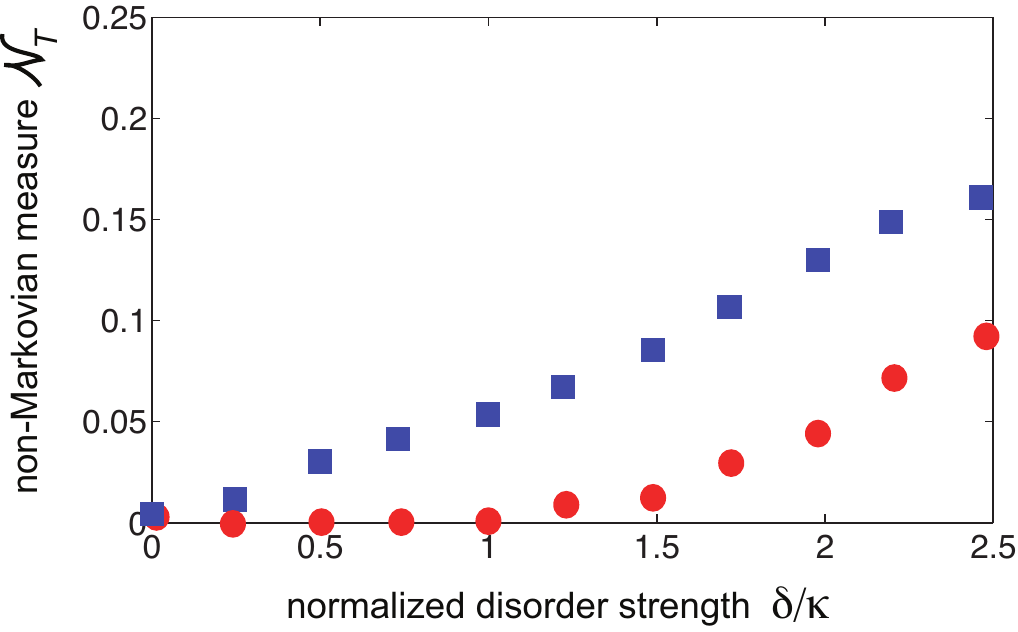}
\caption{{Numerically-computed behavior of the mean value of $\mathcal{N}_T$ versus normalized disorder strength $\delta / \kappa$. Squares and circles refer to the topological trivial ($\varphi=0$) and non-trivial ($\varphi= \pi/2$) phases, respectively. The~mean values have been computed after averaging over 300 realizations of disorder. Other parameter values are $\Delta / \kappa=0.2$ and $\Omega / \kappa=-3/2$.}}
\label{Fig6}
\end{figure}

{As the disorder strength $\delta$ is increased to values larger than $\sim 1-2\kappa$, the~quantum Hall bath becomes an Anderson insulator and chiral edge states are destroyed, i.e., all states become localized~\mbox{\cite{disor1,disor2}}}. Hence, in the strong disorder regime we do not expect a substantial different dynamical behavior, in terms of quantum decay and non-Markovianity, between the $\varphi=0$ and $\varphi= \pi/2$ magnetic fluxes. This~is confirmed by numerical simulations, as~shown in Figures~\ref{Fig5} and \ref{Fig6}. Figure~\ref{Fig5} shows a few examples of decay dynamics and statistical distribution of the non-Markovianity quantifier $\mathcal{N}_T$, over 1000 realizations of disorder, for a disorder strength increased to $\delta=5 \kappa$. Note that in this case strong non-Markovian features are clearly visible both for $\varphi=0$ and $\varphi= \pi/2$, i.e., topological protection of Markovianity in the $\varphi= \pi/2$ case is not anymore observed. Figure~\ref{Fig6} shows the numerically-computed behavior of the mean-value of the non-Markovianity quantifier  $\mathcal{N}_T$ versus disorder strength $\delta / \kappa$, clearly showing the disappearance of topological protection of Markovianity as the disorder strength is increased above $\delta / \kappa \sim 1.25$.

Interestingly, such a result indicates that the transition of the bath, as~the disorder strength is increased,  from a quantum-Hall insulator to an Anderson insulator can be probed by looking at the non-Markovianity of the two-level system dynamics. 

%%%%%%%%%%%%%%%%%%%%%%%%%%%%%%%%%%%%%%%%%%
\section{Conclusions}
In summary we presented a study of the decoherence dynamics of a qubit embedded in a two- dimensional disordered quantum Hall-bath, here embodied by a coupled cavity-array, in the weak- coupling regime, with a synthetic gauge field. Thanks to this tunable synthetic gauge field, it is possible to flip the bath from trivial to non-trivial topological phase, in which the bath spectrum can shows a fractal structure, breaking time-reversal symmetry of the model and leading to appearance of topologically non trivial energy-bands. In particular for rational values of the Peierls' phase the band split into magnetic sub bands with a wide family of topologically protected chiral edge states. 
In the absence of disorder the decoherence dynamics is Markovian in both topological phases, but while in the trivial phase the two-level atom decay arises from the coupling with the bulk states of the bath resonant with the emitter, in the non trivial phase, the~decay arises from the coupling with the chiral edge states of the bath. When disorder is present in the form of random cavity detuning, the~bath exhibits Anderson localization since all of its normal modes are localized. On other hand when the bath is in the non trivial phase, the~chiral edge states result to be robust against localization, protecting the Markovian behavior of system's dynamics. 

Our results provide important physical insights into the fields of topological order and decoherence in open quantum systems, indicating that topological baths with tunable topological properties in the presence of disorder can provide a platform to control decoherence dynamics, either enhancing or suppressing non-Markovian features. 

%%%%%%%%%%%%%%%%%%%%%%%%%%%%%%%%%%%%%%
%%%%%%%%%%%%%%%%%%%%%%%%%%%%%%%%%%%%%%
\acknowledgments{St.L. (Stefano Longhi) and Sa.L. (Salvatore Lorenzo) acknowledge hospitality at IFISC UIB-CSIC, Palma de Mallorca. {G.G. acknowledges financial support from the Maria de Maetzu Program 
for Units of Excellence in R\&D (MDM-2017-0711) and the CAIB postdoctoral program.}}

%%%%%%%%%%%%%%%%%%%%%%%%%%%%%%%%%%%%%%

%%%%%%%%%%%%%%%%%%%%%%%%%%%%%%%%%%%%%%

%%%%%%%%%%%%%%%%%%%%%%%%%%%%%%%%%%%
\end{document}